\documentclass[aps,preprint,fleqn,prb,byrevtex,superscriptaddress,floatfix,citeautoscript]{revtex4}

\usepackage{psfrag,graphicx,longtable}
\newcommand{\ME}[3]{\Bra{#1} #2 \Ket{#3}} 
\newcommand{\Bra}[1]{\left \langle #1 \right |} 
\newcommand{\Ket}[1]{\left | #1 \right \rangle}  
\begin{document}  

\title{Local field effects on the radiative lifetimes of Ce$^{3+}$ in
  different hosts}
\author{Chang-Kui Duan}
\affiliation{
Institute of Modern Physics,
Chongqing University of Post and Telecommunications, Chongqing 400065, China.
}
\affiliation{
Department of Physics and Astronomy, University of Canterbury, Christchurch, New Zealand
}
\author{Michael F. Reid}
\affiliation{
Department of Physics and Astronomy, University of Canterbury, Christchurch, New Zealand
}
\affiliation{MacDiarmid Institute of Advanced Materials and Nanotechnology,
 University of Canterbury, Christchurch,
  New Zealand}

\date{\today}

\begin{abstract}
For emitters embedded in  media of various refractive indices,
 different theoretical models predicted
 substantially different dependencies of the spontaneous emission
 lifetime on refractive
index. It has been claimed that various measurements on $4f\rightarrow
4f$ radiative transition of Eu$^{3+}$ 
in hosts with variable refractive index appear to favor the
real-cavity model [J. Fluoresc. 13, 201 (2003) and references therein,
Phys. Rev. Lett. 91, 203903 (2003)]. We notice that $5d\rightarrow 4f$
radiative transition of rare-earth ions, dominated by allowed
electric-dipole transitions with line strengths less perturbed
by the ligands, serves as a better test of different models.
We analyze the lifetimes of $5d\rightarrow 4f$ transition of 
Ce$^{3+}$ in hosts of refractive indices varying from 1.4 to 2.2.
 The results favor the macroscopic virtual-cavity model based on 
Lorentz local field [J. Fluoresc. 13, 201 (2003)]. 
\end{abstract}

\maketitle

\section{Introduction}

It is well known that radiative transition process of emitters in
media differs from those in
vacuum\cite{Blo1965,Top2003}. Because
of fundamental importance and relevance to various
applications in low-dimensional optical materials and photonic
crystals, this issue continues to attract both theoretical and
experimental attention\cite{Luk2002,Kum2003,Ber2004}. Various
macroscopic (see Ref. \onlinecite{Top2003} for a recent review) and
microscopic \cite{Bor1999,Cre2000,Ber2004} theoretical models have
been developed to predict, among other optical properties, the
spontaneous emission rates of lifetimes on refractive index. 
However, different models predict substantially different 
dependences of radiative lifetime on refractive index. 
The macroscopic model based on Lorentz local field, usually 
referred to as virtual-cavity model\cite{Top2003,Ber2004} has
appeared in most textbooks and been used in calculations.
Only limited experimental studies aimed specifically at discriminating
between different models\cite{Top2003,Kum2003}, with results appear
to support the real-cavity model\cite{Yab1988,Gla1991}. It has also
been pointed out that different models should apply under different 
circumstances\cite{Top2003}.

The underline assumption of all those models and experimental
 studies is that the only contribution to the spontaneous
radiative lifetime is from the electric dipole moment whose strength
does not vary ( or changes in a predictable way) when surrounding media
vary. We notice that the experimental results that have been claimed to 
support the real-cavity model\cite{Rik1995,Sch1998,Kum2003} 
are all lifetimes of the $^5D_0$ level of Eu$^{3+}$ in different hosts
with varying refractive index. It is well-known that
part of the radiative relaxation of $^5D_0$ (to $^7F_1$) is due to
 magnetic dipole moment, which has a different dependence on 
refractive index, and the electric dipole strength of $^5D_0$ to
 $^7F_2$ transition is hypersensitive to environment and may not
be treated as a constant. In general, $4f\rightarrow 4f$ electric
 dipole radiative relaxation
in rare-earth ions is due to mixing in $4f^N$ states with states 
with opposite parity, which depend strongly on the environment. Since
this dependence is usually very difficult to be taken into account,
lifetimes of $4f\rightarrow 4f$ radiative relaxation do not serve as
a good examination of different models. In contrast, $5d\rightarrow 4f$
radiative transitions of rare-earth ions are dominated by
 allowed electric-dipole moment contributions, whose strengths are
 less perturbed by the environment and the line strengths for the radiative
 relaxation can be reliably  predicted. Hence the lifetimes of
$5d\rightarrow 4f$ radiative transitions give a better test of 
different models. 

In this paper we analyze the lifetimes of $5d\rightarrow 4f$ transition
of Ce$^{3+}$ ions in hosts of different refractive indices and make
a comparison between different models. In Sec. II we derive the basic
formula to calculate the line strength and lifetime of the $d$ levels
of Ce$^{3+}$. The lifetimes and energies of $d$ levels of Ce$^{3+}$ 
ions and the refractive indices are summarized and analyzed with
different models in Sec. III. 

\section{$d\rightarrow f$ transition rates  of Ce$^{3+}$ in
  hosts}
The general spontaneous radiative emission rate of
 electric dipole transition from an localized initial state $I$ to
a localized final state $F$ can be written as \cite{Top2003}
\begin{equation}
\Gamma_{IF}= \frac{64\pi^4}{3h} \chi \nu_{IF}^3 |\vec {\mu}_{IF}|^2,
\label{ind_rate}
\end{equation}
where $I$ and $F$ and transition initial and final states,
respectively, $\nu_{IF}$ is the emission wavenumber, 
$\vec{\mu_{IF}}$ is the electric dipole moment $-e\vec{r}$ 
between state $I$ and
$F$, and $\chi$ is an enhancement factor due to dielectric medium,
which equals $n[(n^2+2)/3]^2$ for virtual- and $n[3n^2/(2n^2+1)]^2$
for real-cavity model. The lifetime of energy $I$ can be calculated
as the inverse of the total emission rate of $I$. 

For the $5d\rightarrow 4f$ emission of Ce$^{3+}$ ions, 
The eigenvectors of transition initial states are dominated by 
bases with only one electron in open shell $5d$ and the transition
final states are dominated by bases with only one electron in
open shell $4f$. it is tempting
to approximate the electric dipole moment between a $5d$ state and
a $4f$ state with the straightforward matrix element of electric
dipole $-e\vec{r}$ between one particle orbitals $4f$ and
$5d$.  Such an approximation overestimate the radiative lifetime of
Ce$^{3+}$ free ion by a factor of about 3. 
Since the transition initial and final states are actually
many-particle states, calculation \cite{Zha2001}
showed that configuration mixing needed to be taken into account
to obtain correct radiative lifetime for Ce$^{3+}$ free ion.
For rare-earth ions in hosts, ligand polarization could also contribute
to the radiative transition rate. Theoretical treatment of 
$f-d$ electric dipole moment of rare-earth ions taking all those
corrections into account is not trivial, which can be found
in Ref.\onlinecite{Dua2005a}.  For Ce$^{3+}$ ions in 
hosts, since there is only one electron in the open shell,
neglecting the small ligand polarization contribution,
 the correction due to configuration mixing is equivalent to
reduce the radial integral $\ME {5d} r {4f}$. For Ce$^{3+}$ free ion,
the effective radial integer is $\ME {5d} r {4f}_{\rm eff} = 0.025$nm.

For Ce$^{3+}$ ions, since the splitting between different transition
final states  is much smaller than the average energy difference 
between the lowest $5d$ and $4f$ states, we can make an approximation
to the summation of  Eq.\ (\ref{ind_rate}) over final state $F$
 by replace the wave numbers
with the average value $\bar{\nu}$. Under this approximation, the total 
spontaneous emission rate turns out to be independent of
the wavefunction of the initial $5d$ state, and can be written as
\begin{eqnarray}
\label{theory}
\frac{1}{\tau_r} &=& \frac{64 \pi^4 e^2 \chi |\ME {5d} {r}
  {4f}_{\rm eff}|^2\bar{\nu}^3}{5h}
\\
 &=& 4.34 \times 10^{-4} |\ME {5d} {r} {4f}_{\rm eff}|^2 
    \chi \bar{\nu}^3 (s^{-1}),
\end{eqnarray}
where units for radial integral, $\bar{\nu}$ and $\tau_r$ 
are nm, cm$^{-1}$ and sec, respectively. With measured $\tau_r$ and 
$\bar{\nu}$ values, we can derive measured values for
$\ME {5d} r {4f}^2 \chi$  ($\sim \tau_r^{-1}$) and compare them
 with the predictions of different model.

\section{Analysis of radiative relaxation lifetimes of Ce$^{3+}$
  in different hosts}

The $5d\rightarrow 4f$ transitions of Ce$^{3+}$ in various hosts
have been widely studied due to applications as scintillators, tunable
UV lasers and phosphors. The lifetimes, peak wavelengths of emission
spectra and refractive indices of Ce$^{3+}$ in different hosts are
summarized in Table \ref{table1}. Some of the data are measured at
room temperature and some are measured at low temperature.
Ideally, we need work with the lifetimes for different hosts
 at the same low temperature, preferably at 0K. Fortunately,
 due to large separation between $5d$ and $4f$ states and strong
 electric dipole $5d-4f$  radiative relaxation, nonradiative
 relaxations are negligible and the lifetimes at room temperature
 only change (decrease) slightly from low-temperature ones. 
In some experiments, the observed lifetimes
at room temperature is even slightly longer than the low-temperature
lifetimes due to reabsorption. We neglect all these small corrections 
and put a uncertainty of about $10\%$ to the spontenous emission
lifetime in the figure to guide eyes.

\begin{figure}[htp]
\includegraphics[width = 14cm]{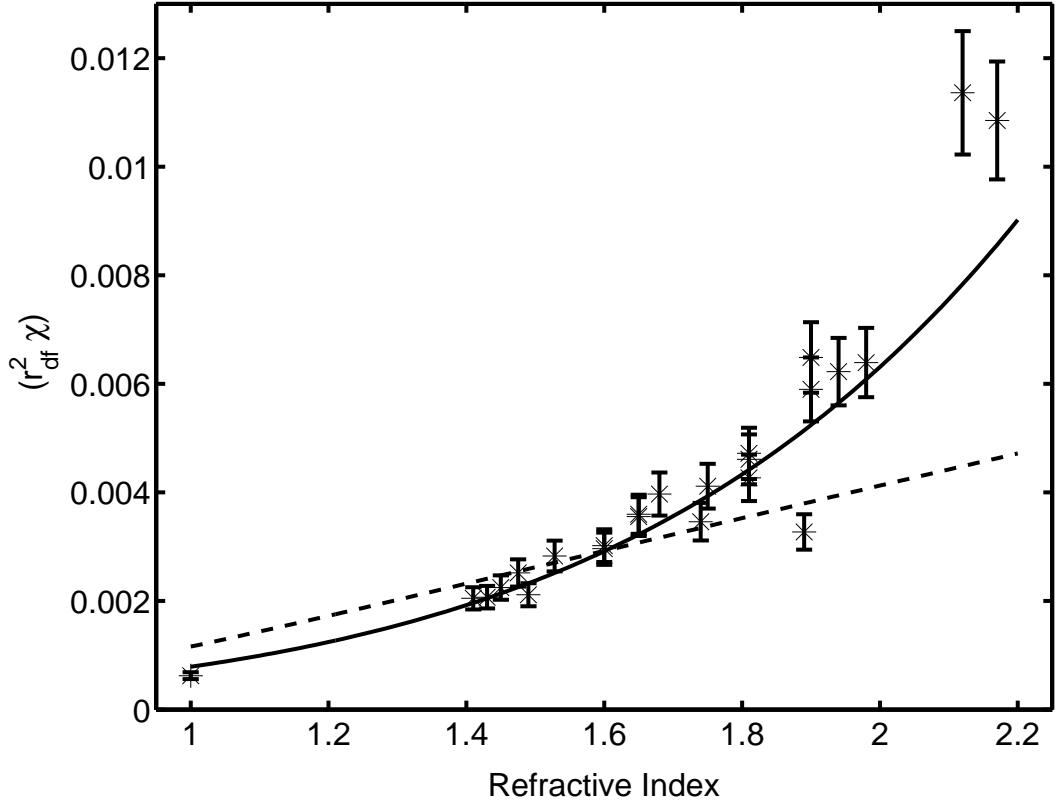}
\caption{
Variation of $(\ME {5d} r {4f}_{\rm eff} \chi)$ with refractive
index. The experimental values are plotted as '*' with a $10\%$ error
bar to guide eyes. The solid curve is calculated with virtual-cavity model using
best least-square-fitting value $\ME {5d} r {4f}_{\rm eff} =
0.0281$,  and the dashed curve is for real-cavity model with 
$\ME {5d} r {4f}_{\rm eff}^{\prime} = 0.0341$.
\label{figure}
}
\end{figure}

Since the transition rates depend not only on refractive index factors but
also the emission energy, we cannot follow Ref.s \ ~ \onlinecite{Top2003,Kum2003}
to compare experimental and theoretical lifetime-refractive index curves.
Instead, the measured $\ME {5d} r {4f} ^2 \chi$ values are plot as a
function of measured refractive index in Fig.\ref{figure}, together
with calculated curves using two different models with 
$\ME {5d} r {4f}_{\rm  eff}^2$ values obtained with
experimental-value-weighted least-square
fitting. It can be seen that the virtual-cavity model fits the measured
data much better than the real-cavity model, while the real-cavity
model gives an almost linear dependence of the emission rates on refractive
index, which cannot fit the measured data at all. This is in 
contrary to the conclusion draw from the $f-f$ transitions of
 Eu$^{3+}$ in various
hosts. The best-fit value for the effective radial integral is 
$\ME {5d} r {4f}_{\rm eff} = 0.0281$. This value is actually bigger
than the free ion value $0.025$, in contrary to expectations
that it should be smaller than the free ion
value\cite{Kru1966,Lyu1991}.
Using the virtual-cavity model, the $\ME {5d} r {4f}_{\rm eff}$ for
each hosts have been calculated and are given in Table \ref{table1}. 
It can be seen that most of the values are quite consistent.
\section{Conclusion}
In conclusion, we analyze the spontaneous emission rates  $5d\rightarrow 4f$
 transition of Ce$^{3+}$ in hosts of refractive indices between 1.4 to 2.2 with
the two major models. The dependence of the rates on refractive indices
 favor the macroscopic  virtual-cavity model based on Lorentz local field 
\cite{Top2003}. We also conclude that the 
values of Ce$^{3+}$ effective radial integral
 $\ME {5d} {r} {4f}_{\rm eff}$ are larger in crystals than in vacuum.

\section*{Acknowledgment}

C.K.D. acknowledges support of this work by the National Natural Science
Foundation of China, Grant No. 10404040 and 10474092.

\bibliography{lifetime,celifetime}

\squeezetable
\begin{table}[htp]
\caption{\label{table1}
Summary of the radiative decay parameters for Ce$^{3+}$ in various
hosts,
where $\tau_{\rm r}$ (unit: ns) is the measured lifetime of the lowest
$5d$ state, which is dominated by radiative relaxation and
used an spontaneous emission lifetime in this paper,
 $\lambda$ (unit: nm) is the peak emission wavelength, $n$ is refractive index,
$\chi_{\rm virtual}$ and $\chi_{\rm real}$ are $\chi$-factors
for virtual- and real-cavity models, respectively, and $\ME {4f} r
{5d} _{\rm eff}$ (unit: nm) is derived from measured lifetime using
Eq. (\ref{theory}) for virtual-cavity model.
}
\begin{ruledtabular}[htp]
\begin{tabular}{llllllll}
Host            &Ref.       &$\tau_{\rm r}$&$\lambda$    &
$n$       &$\chi_{\rm virtual}$&$\chi_{\rm real}$& $\ME{4f}r{5d}$ \\\hline
LaF$_3$         &\onlinecite{Lyu1991} &
 19  & 292 &   1.6  & 3.69 &     2.52 &     0.0286\\
LaF$_3$         &\onlinecite{Ped1992} &
 21 &  300 &   1.6 &  3.69  &    2.52   &   0.0283\\
YAG             &\onlinecite{Lyu1991} &
 59.1 &550 &   1.9 &  6.64  &    3.30   &   0.0312\\
YAG             &\onlinecite{Ham1989} &
 65   &550 &   1.9 &  6.64  &    3.30   &   0.0298\\
CaF$_2$         &\onlinecite{Mir1996} &
 40   &330 &   1.43&  2.59  &    2.07  &    0.0282\\
YAlO$_3$        &\onlinecite{Lyu1991} &
 17.1& 362  &  1.98&  7.71  &    3.50   &   0.0288\\
YLiF$_4$        &\onlinecite{Lyu1991} & 
 35.7& 320 &   1.49&  2.94  &    2.23  &    0.0268\\
Gd$_2$SiO$_5$   &\onlinecite{Pid2003} & 
 56 &  430 &   1.89&  6.52  &    3.27  &    0.0224\\
Lu$_2$SiO$_5$   &\onlinecite{Pid2003} & 
 40 &  420  &  1.81 & 5.59  &    3.06 &     0.0276\\
Lu$_2$SiO$_5$   &\onlinecite{Suz1993} &
 32 &  400  &  1.81&  5.59  &    3.06   &   0.0287\\
Lu$_2$SiO$_5$   &\onlinecite{Suz1993} &
 54 &  480  &  1.81 & 5.59  &    3.06  &    0.0290\\
LuAlO$_3$       &\onlinecite{Pid2003} &
 18 &  365 &   1.94 & 7.16 &     3.40   &   0.0295\\
Lu$_2$Si$_2$O$_7$&\onlinecite{Pid2003}& 
 38 &  385  &  1.74 & 4.88   &   2.88  &    0.0266\\
Li-Al-B glass   &\onlinecite{Das1998} & 
 38 &  360  &  1.528& 3.19  &    2.33  &    0.0298\\
Sr$_2$B$_5$O$_9$Br&\onlinecite{Dot1999}&
 38 &  390  &  1.65&  4.08   &   2.64  &    0.0297\\
Sr$_2$B$_5$O$_9$Br&\onlinecite{Dot1999}& 
 29 &  355  &  1.65&  4.08  &    2.65  &    0.0295\\
LiSrAlF$_6$     &\onlinecite{Mar1994} &
 28 &  292 &   1.41 & 2.49    &  2.02   &   0.0287\\
LiCaAlF$_6$     &\onlinecite{Mar1994} &
 25 &  290   & 1.45 & 2.71    &  2.13  &    0.0288\\
CaS             &\onlinecite{Hos1980} &
 36 &  562  &  2.12 & 9.93  &    3.86  &    0.0338\\
SrGa$_2$S$_4$   &\onlinecite{Hos1980} & 
 20 &  455 &   2.17&  10.8  &    3.99   &   0.0316\\
BaF$_2$         &\onlinecite{Woj2000} & 
 30 &  320 &   1.475& 2.85  &    2.19   &   0.0297\\
Ca$_2$Al$_2$SiO$_7$&\onlinecite{Yam2002}& 
 40 &  410&    1.68 & 4.34&      2.73 &     0.0302\\
YPO$_4$         &\onlinecite{Lar2001} & 
 23  & 345 &   1.75&  4.98&      2.91  &    0.0287\\
Free ion        &\onlinecite{Zha2001} & 
 30  & 201  &  1  &   1     &    1    &     0.0250\\
\end{tabular}
\end{ruledtabular}
\end{table}

\end{document}